\documentclass[letterpaper,english,reprint,aps,fleqn]{revtex4-1}
\usepackage[latin9]{inputenc}
\usepackage{amsmath}
\usepackage{amssymb}
\usepackage{graphicx}

\makeatletter

\usepackage{babel}

\usepackage{braket}

\usepackage{babel}

\makeatother

\usepackage{babel}
\begin{document}

\preprint{APS/123-QED}

\title{Quantum Entanglement in Time for a Distributed Ledger}

\author{Nils Paz }

\affiliation{ZapLinQ , San Diego, California}

\author{Steven Silverman}

\affiliation{Cal State University San Marcos , San Marcos, California}

\author{John Harmon }

\affiliation{Independent , San Diego, California}

\date{\today}
\begin{abstract}
Distributed Ledger Technology (DLT) is a shared, synchronized and
replicated data spread spatially and temporally with no centralized
administration and/or storage. Each node has a complete and identical
set of records. All participants contribute to building and maintaining
the distributed ledger. Current DLT technologies fall into two broad
categories. Those that use block-chains such as in Bitcoin or Ethereum,
and newer approaches which reduce computational loads for verification.
All current approaches though difficult to crack can be vulnerable
to quantum algorithms using Quantum Information Technologies (QIT).
This effort joins the 2 technologies, constructing a Quantum Distributed
Ledger (QDL) which provides a higher level of security using QIT and
a decentralized data depository using DLT. This enhanced security
prevents middleman attacks with quantum computers yet retains the
advantages of a decentralized ledger of data.
\begin{description}
\item [{GCAPS~numbers}] 05.20.Dd, 05.70.Ln, 05.70.Np, 88.05.De.{\small \par}
\end{description}
\end{abstract}


\maketitle

\section{\label{sec:level1}Introduction:}

Current DLT implementations, whether using a \textquotedblleft block
chain\textquotedblright{} or a Directed Acyclic Graph (DAG) approach
are limited by their use of a classical computing system to encrypt,
decrypt and verify. In the block chain\citep{nakamoto2012bitcoin},
it is further limited by its intensive computational load as the proof
of work of the block chain keeps increasing as the chain lengthens
over time. The DAG approach to DLT, though removing the effort of
the proof of work to the entire chain of events, still implements
classical computing system algorithms to verify chain snippets. The
classical approach is vulnerable to quantum computational attacks.
Prevention of any middleman attack can be thwarted by using quantum
entanglement. Quantum entanglement is one the major attributes of
Quantum Information Technology (QIT). It is also known as Einstein,
Podolsky, Rosen (EPR)\citep{PhysRev.47.777} pairing. The techniques
and validation of quantum entangled processes are now validated and
successfully implemented in commercial QIT such as quantum computing.

Quantum entanglement or EPR pairing produces a physical phenomenon
where a pair of fields or particles behave as if they were the same
field or particle both in space and time. An example of this would
be if 2 separate points in space (Alpha and Beta) cooperate to entangle
a pair of photons. Alpha measures photon A and Beta measures photon
B. Alpha\textquoteright s individual measurements of A appear as random
fluctuations. The same is true for Beta\textquoteright s measurements
of B. However, if they compare the results of these 2 entangled particles
they are either 100\% correlated or 100\% inversely correlated. This
phenomenon is quantum entanglement.

This research relates to the concept of a Quantum Block-chain\citep{Tessler2017}
which we discuss within the context of Distributed Ledger Technology\citep{Walport2015}
systems for many applications such as: currency transactions, banking,
healthcare, logistics, business and personal contracts, chain of custody
and more. We use the synchronization of DLT systems using quantum
entanglement in time. This state can only be achieved using quantum
entanglement or Einstein-Podolsky-Rosen (EPR) pairing. This physical
state pairs a field or a field of particles which results in a behavior
as if they were the same field or particle in both space and time,
though separated by large distances. The paired entities would share
states, and either be inversely correlated or directly correlated.
In addition, if the entangled entities are placed through the process
of a quantum teleportation state through time, the state produced
is correlated with a no longer existent entity entangled in the past.
Using this phenomenon of Quantum Entanglement In Time (QuEIT)\citep{Megidish2013},
the originating source and the data coherent with the entangled source
would be immune to cryptographic attacks, retain the fidelity of the
chain of events and be impervious to the deletion of any set of data
in the string of the distributed ledger.

\includegraphics[width=8cm,height=3cm]{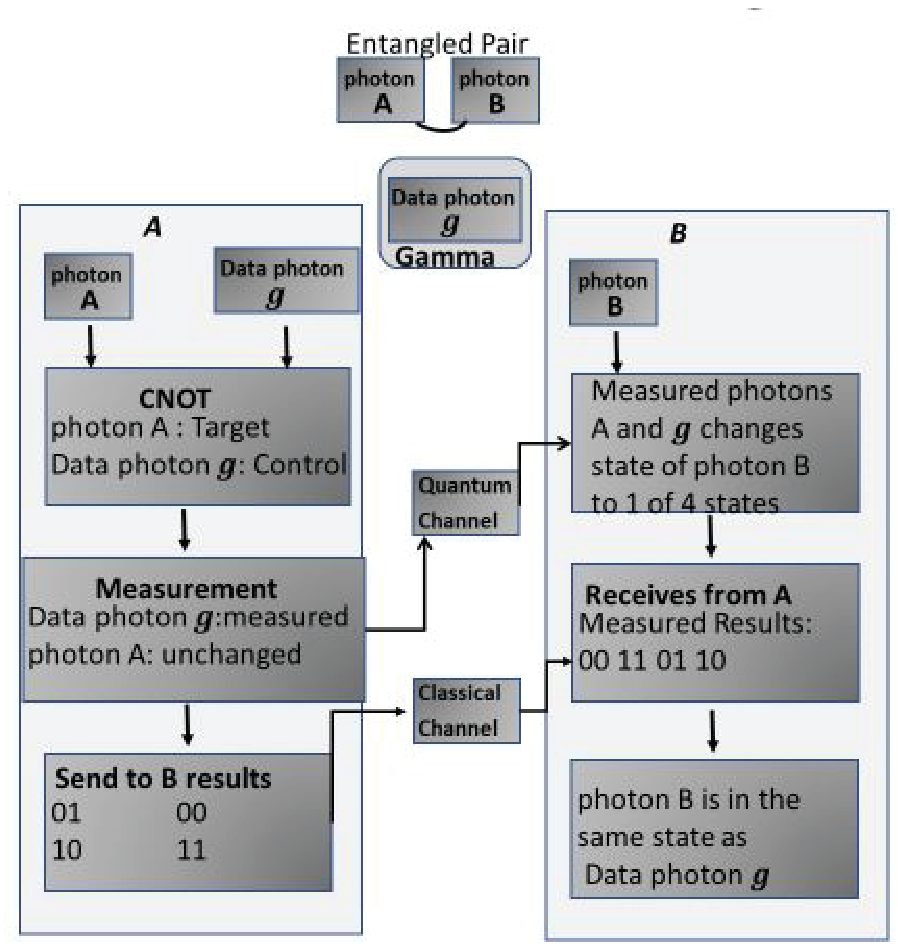}

Figure 1: Quantum Entanglement as it relates to QDL or Quantum Block
chains

The fundamental basis of a Quantum Block-chain is the quantum teleportation
schema\citep{Marinescu2004}. See Figure 2 , where \textbar{} g \textgreater{}
 represents the block-chain. In this format we see the fundamental
role played by the standard design of quantum teleportation.

\includegraphics[width=7cm,height=7cm]{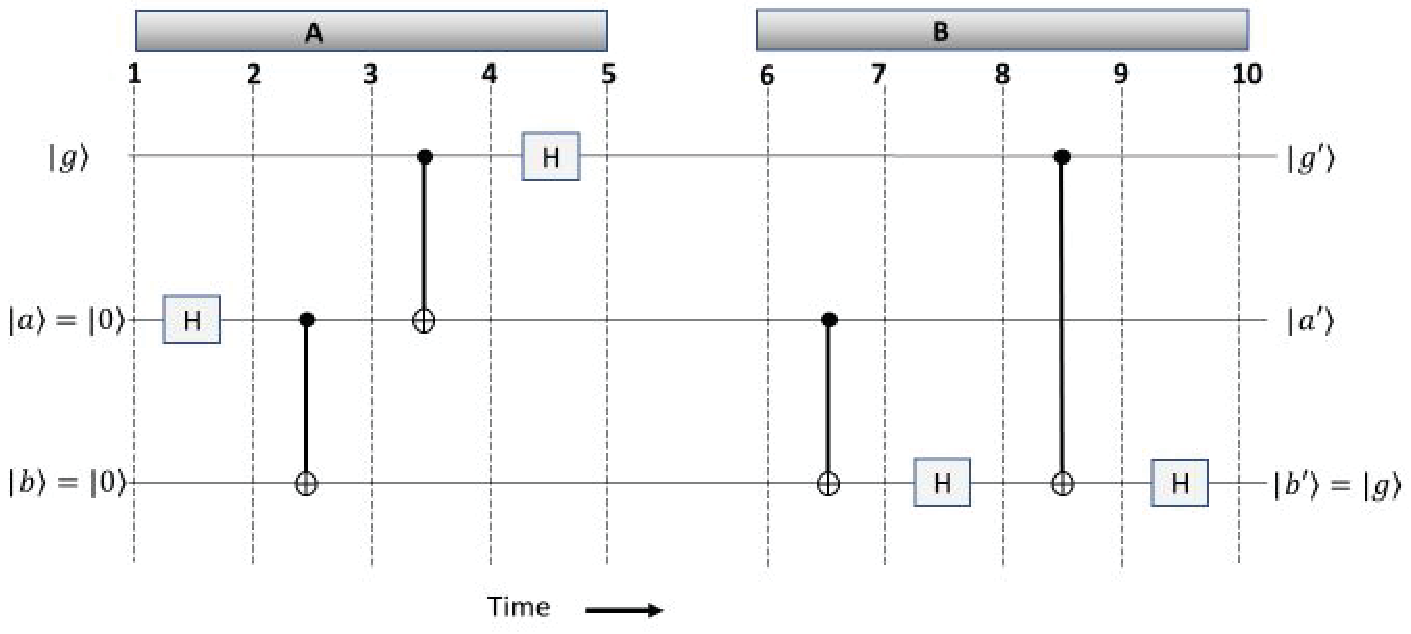}

Figure 2: Basic circuit for Quantum Entanglement in Time ( QuEIT )
in a Distributed Ledger 

\section{Mathematical Developments}

The fundamental basis of a Quantum Block-chain is the quantum teleportation
schema. See figure 2 , where \textbar{} g \textgreater{}  represents
the block-chain. Following, Marinescu\citep{Marinescu2004}, we form
the input vector as \textbar{}g\textgreater{}$\otimes$$|h>\otimes|j>,$
(tensor\textendash exterior product ),  which is represented as a
8 x 1 vector in Hilbert Space. as in equation (1).

\begin{equation}
\left(\begin{array}{c}
g_{1}h_{1}\text{j}_{1}\\
g_{1}h_{1}\text{j}_{2}\\
g_{1}h_{2}\text{j}_{1}\\
g_{1}h_{2}\text{j}_{2}\\
g_{2}h_{1}\text{j}_{1}\\
g_{2}h_{1}\text{j}_{2}\\
g_{2}h_{2}\text{j}_{1}\\
g_{2}h_{2}\text{j}_{2}
\end{array}\right)
\end{equation}

where the basic ket is defined as:
\begin{equation}
|g>=\left(\begin{array}{c}
\text{\ensuremath{g_{1}}}\\
\text{\text{\ensuremath{g_{2}}}}
\end{array}\right)
\end{equation}

There are 10 stages shown in Figure 2. Stage one is the initial input
, equation 1. The output vector from stage 1 would have the mathematical
form:
\begin{equation}
|stage\text{1}>=\left(\begin{array}{c}
g_{1}h_{1}\text{j}_{1}\\
g_{1}h_{1}\text{j}_{2}\\
g_{1}h_{2}\text{j}_{1}\\
g_{1}h_{2}\text{j}_{2}\\
g_{2}h_{1}\text{j}_{1}\\
g_{2}h_{1}\text{j}_{2}\\
g_{2}h_{2}\text{j}_{1}\\
g_{2}h_{2}\text{j}_{2}
\end{array}\right)
\end{equation}
\textbar{}g\textgreater{} $\otimes|h>\otimes|j>,$ (8 x 1 tensor)

Then the output vector \textbar{} stage 2\textgreater{} is expressed
as: $\otimes$

\begin{equation}
|stage2>=\left(\begin{array}{cc}
1 & 0\\
0 & 1
\end{array}\right)\otimes\frac{1}{\sqrt{2}}\left(\begin{array}{cc}
1 & 1\\
1 & -1
\end{array}\right)\otimes\left(\begin{array}{cc}
1 & 0\\
0 & 1
\end{array}\right)\otimes|stage1>
\end{equation}

The middle matrix in equation 4 is the Hadamard Matrix , while the
left and right matrices are the Identity Matrices. This concatenation
continues until stage 10 where the final output vector \textbar{}
stage 10\textgreater{} is represented as:
\begin{equation}
|stage10>=\left(\begin{array}{cc}
1 & 0\\
0 & 1
\end{array}\right)\otimes\frac{}{}\left(\begin{array}{cc}
1 & 0\\
0 & 1
\end{array}\right)\otimes\frac{1}{\sqrt{2}}\left(\begin{array}{cc}
1 & 1\\
1 & -1
\end{array}\right)\otimes|stage9>
\end{equation}

The details of this process are left in the Appendix. The final vector
stage10 involves 30 matrices and nine 8 x 1 vectors, for a total of
39 matrices. This for only a three channel ( 3 qubit ) input. The
stage 10 vector is factored into the following:
\begin{equation}
|stage10>=\left(\begin{array}{c}
g_{1}\\
g_{2}
\end{array}\right)\otimes(entangled\ states\ of\ |b>|c>)
\end{equation}

It should be observed that the input to the top channel \textbar{}g\textgreater{}
which represents the Block-chain ( or distributed ledger ) appears
teleported to the remote bottom channel. This event ( position and
time ) In this simplified arrangement, the Block-chain represents
4 classical bits , $2^{2}.$In our design a terabyte of information
would require a 12 qubit object, for 2$^{12}$ classical bits. That
is $|g>=|g_{first}>\otimes|g_{second}>...\otimes|g_{fortyith}>,$each
\textbar{}g$_{i}>$being a 2 x 1 vector. According to Lloyd\citep{Gell-Mann1996}
the order of complexity for a terabyte N=10$^{12}$ follows the logarithmic
relationship $\sim$ log$_{2}N$ which is essentially 40 ( qubits
needed for encoding ). The summary of this process is shown in Figure
3 below. 

\includegraphics[width=7cm,height=7cm]{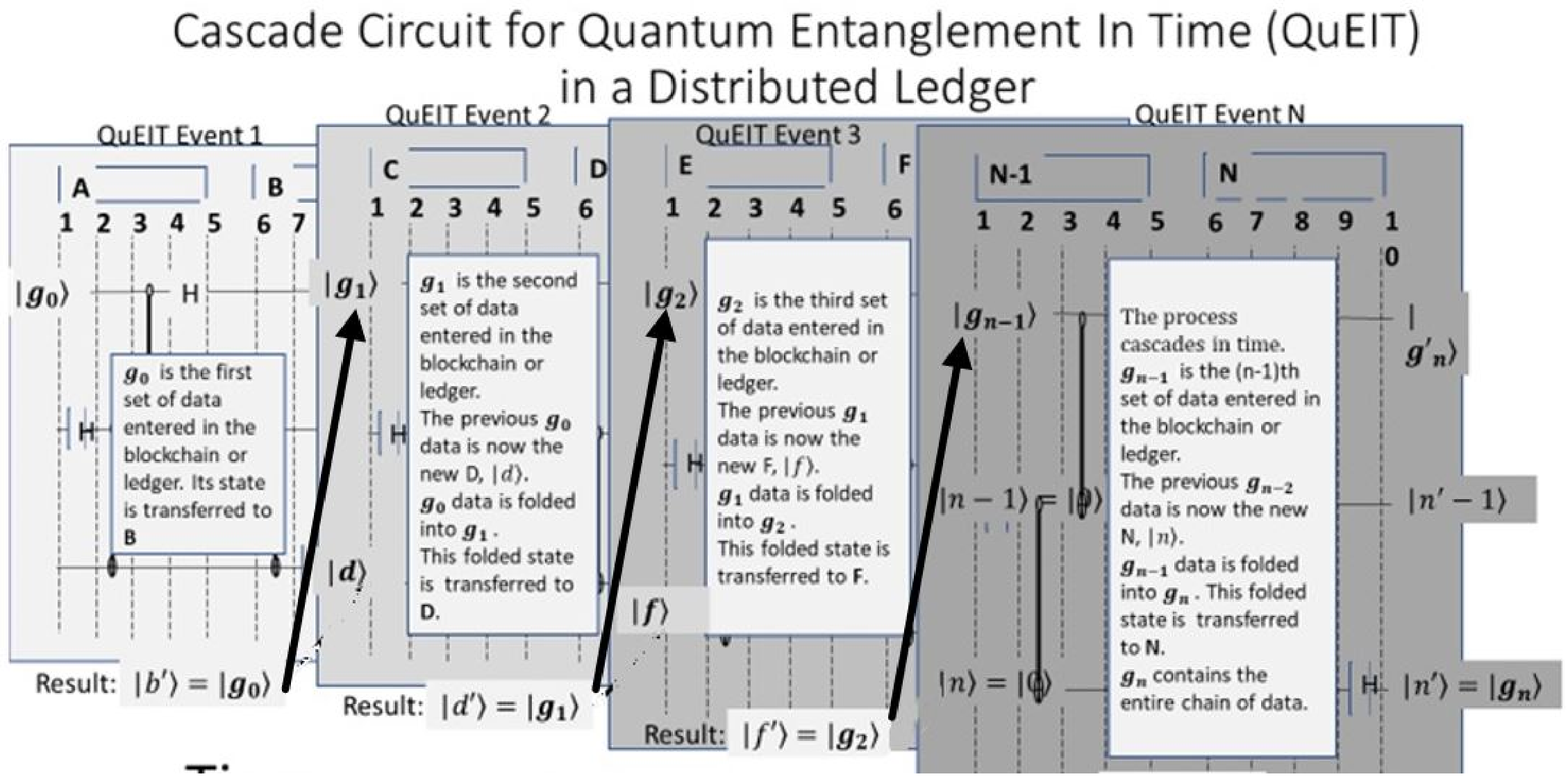}

Figure 3: Cascade of the basic circuit for QDL

Given this architectural design the quantum mechanical correctness
of teleportation can not be denied. Schrodinger's or Heisenberg's
equations will dynamically exhibit the solution as described in Figure
1. However to actually implement this circuit schema is another matter.
Two major events have to exist for the Block-chain to be teleported.
One is the preparation of the initial state vector called \textbar{}g\textgreater{}
here. Both end points, sender and receiver ( referred to in nearly
all descriptions as Alice at stage 1 and Bob at stage 2 ) need to
share an entangled ket. This could be prferred with another dual channel
what would be called in the AT\&T\citep{Baxter1985} long distance
or local distance loop as a setup or control channel. This separate
channel is just for set up only, while then the ''data'' channel
would be the channel for data transport only. Most texts consider
this control channel as classical since a quantum channel would not
in principal be needed. However, there is no reason for this. Dual
quantum channels is the future of telecom anyway whether they be dual
virtually ( single quantum channel implemented ) or a dual physical
channel. The second point not addressed in the circuit diagram is
the measurement process. The remote ( Bob ) needs to collapse some
of the transported Block-chain to recover and add to the Block. This
is the purpose of the Block-chain ( ledger ) to add securely to the
chain and move on. In some designs, see Chuang and Nielsen\citep{nielsen00},
the measurement is shown as part of the circuit. The authors here
feel that this is overloading and adding some confusion to the flow.
Instead we will incorporate the measurement in a more appropriate
graphic that shows a possible implementation of the Figure 1 as a
network of photons and beam splitters. Since, in this design a measurement
is more naturally rendered visually.

\section{Implementation of Teleportation Schema for a Distributed Ledger}

In this section we show an implementation of the Block-chain teleportation
scheme using beam splitters and photons. The implementation of Hadamard
and CNOT gates is reported in the literature\citep{Marinescu2004}.
Some words are in order for the description of the DLT and its encoding
and teleportation. Several more steps are needed to complete the insertion
of data, its teleportation and its measurement through the mining
process. Looking at figure 2, the top state, on input is part of an
initially entangled pair of states. They are shared between two entities.
Sender and Receiver. This will involve a measurement process on initialization.
The blockchain is represented as the lowest ket on the left side.
Now there is some added data to the distributed ledger ( top left
side of figure 2 ). Part of which is the control and some part the
target qubits. Particles ( photons A ) and Data Photons B are initially
in a maximally entangled state. This state can be respresented in
our analysis as a terabyte of data that will require up to 40 qubits
( photon states ). Now the third party to whom the Blockchain is to
be transported to has photons B. Photons A and Photons B are maximally
entangled. The simple picture shown for illustrative purposes shows
the entangled states as two qubits. In actuality they are are tensor
product of as many as 40 qubits. Now after this has been effected,
we are ready to proceed as input vector to Figure 3. Figure 2, as
shown here, is formally correct but lacks the measurement steps and
also the initial entanglements as well. This was only done for reasons
of clarity. The diagram would be too busy if the measurements were
included in Figure 3, so they are omitted.

\section{Results and Summary}

In Section II, we described the mathematics necessary to model the
implementation of this Quantum Distributed Ledger (QDL). The mathematics
for QDL is an expansion of the mathematics, principles and experimentation
well studied and implementated for over 15 years. Our mathematical
expressions expand the concepts born out both experimental implementation
and theoretical prediction. Section III provides the outline for the
implemetation of QDL. It is coherent and robust given that its foundation
is based on almost two decades of experimental results worldwide.
The circuit diagram provides a prototype for future exploration and
current implementation.

The Quantum Distributed Ledger is a robust implementation of both
technologies using DLT and QIT. Using Quantum Entanglement In Time
(QuIET), its implementation extends quantum teleportation to support
movement of data through space and across time. The initial Qbit is
the seed which builds and aggregates data thereby creating a chain
of information to form a quantum distributed ledger that is secure,
distributed and extendable.

\appendix

\end{document}